\documentclass[letterpaper, 10 pt, conference]{ieeeconf}  

\IEEEoverridecommandlockouts                              

\usepackage{amsmath} 
\usepackage{amssymb}  
\usepackage{cite}
\usepackage{graphicx}
\usepackage[caption=false, font=footnotesize]{subfig}
\usepackage{placeins}
\usepackage[utf8]{inputenc}
\usepackage{tikz}

\newcommand\copyrighttext{%
	\footnotesize Copyright $\copyright$ 2019 IEEE.
	Personal use of this material is permitted.
	Permission from IEEE must be obtained for all other uses, in any current or future media, including reprinting/republishing this material for advertising or promotional purposes, creating new collective works, for resale or redistribution to servers or lists, or reuse of any copyrighted component of this work in other works.}%
\newcommand\copyrightnotice{%
	\begin{tikzpicture}[remember picture,overlay]%
	\node[anchor=south,yshift=10pt] at (current page.south) {\fbox{\parbox{\dimexpr\textwidth-2cm}{\copyrighttext}}};%
	\end{tikzpicture}%
	\vspace{-10pt}%
}

\title{\LARGE \bf
A Subjective-Logic-based Reliability Estimation Mechanism for Cooperative Information with Application to IV's Safety$^\ast$
}

\author{Johannes M{\"u}ller$^{1}$, Michael Gabb$^{2}$, and Michael Buchholz$^{1}$
	\thanks{*This work was financially supported by the Federal Ministry of Economic Affairs and Energy of Germany within the program "Highly and Fully Automated Driving in Demanding Driving Situations" (project MEC-View, grant number 19A16010I).}
	\thanks{$^{1}$Johannes M{\"u}ller and Michael Buchholz are with the Institute of Measurement, Control and Microtechnology,
		Ulm University, D-89081 Ulm, Germany
		{\tt\small \{johannes-christian.mueller, michael.buchholz\}@uni-ulm.de}
	}%
	\thanks{$^{2}$Michael Gabb is with the Robert Bosch GmbH, 74003 Heilbronn, Germany
		{\tt\small michael.gabb@de.bosch.com}}%
}

\begin{document}

\maketitle
\copyrightnotice%
\thispagestyle{empty}
\pagestyle{empty}

\begin{abstract}

Use of cooperative information, distributed by road-side units, offers large potential for intelligent vehicles (IVs). As vehicle automation progresses and cooperative perception is used to fill the blind spots of onboard sensors, the question of reliability of the data becomes increasingly important in safety considerations (SOTIF, Safety of the Intended Functionality).

This paper addresses the problem to estimate the reliability of cooperative information for in-vehicle use. We propose a novel method to infer the reliability of received data based on the theory of Subjective Logic (SL). Using SL, we fuse multiple information sources, which individually only provide mild cues of the reliability, into a holistic estimate, which is statistically sound through an end-to-end modeling within the theory of SL.

Using the proposed scheme for probabilistic SL-based fusion, IVs are able to separate faulty from correct data samples with a large margin of safety. Real world experiments show the applicability and effectiveness of our approach.

\end{abstract}

\section{INTRODUCTION}

Being already widely adopted in the avionics industry and navigation \cite{pullen2011}, monitoring and assuring systems' functional performance based on multiple error sources plays an increasingly important role for intelligent vehicles (IVs) and is generally termed \emph{Safety of the Intended Functionality} (SOTIF) in the automotive context. 
Cooperative information can increase the efficiency of motion planning for IVs. Relying on wrong information, however, may lead to emergency braking or even crashes. In order to reach SOTIF, measures have to be taken to avoid such incidences.
Consider, for example, an IV approaching an intersection with occlusion of the crossing road for on-board sensors. In order to resolve the occlusion, infrastructure sensor modules are posted at the intersection. From the sensor data, an environment model is computed. This model then is reported to the IV by a road side unit (RSU). 

While classical probabilistic approaches can easily model uncertainty in the object states, often described by covariance matrices, they lack the ability to explicitly model the evidence-based statistical uncertainty of the estimated probabilities \cite{Joesang2016}.
In the example mentioned above, it can be intuitively seen that 
the reliability of the incoming data depends on the amount of evidence supporting them. While classical probabilistic approaches can easily model the probability of the hypothesis that a road user is at a given position at a given time, they lack the ability to express how much evidence actually supports the hypothesis and what is the influence of the assumed prior.

\begin{figure}
	\centering
	\includegraphics[width=0.92\linewidth]{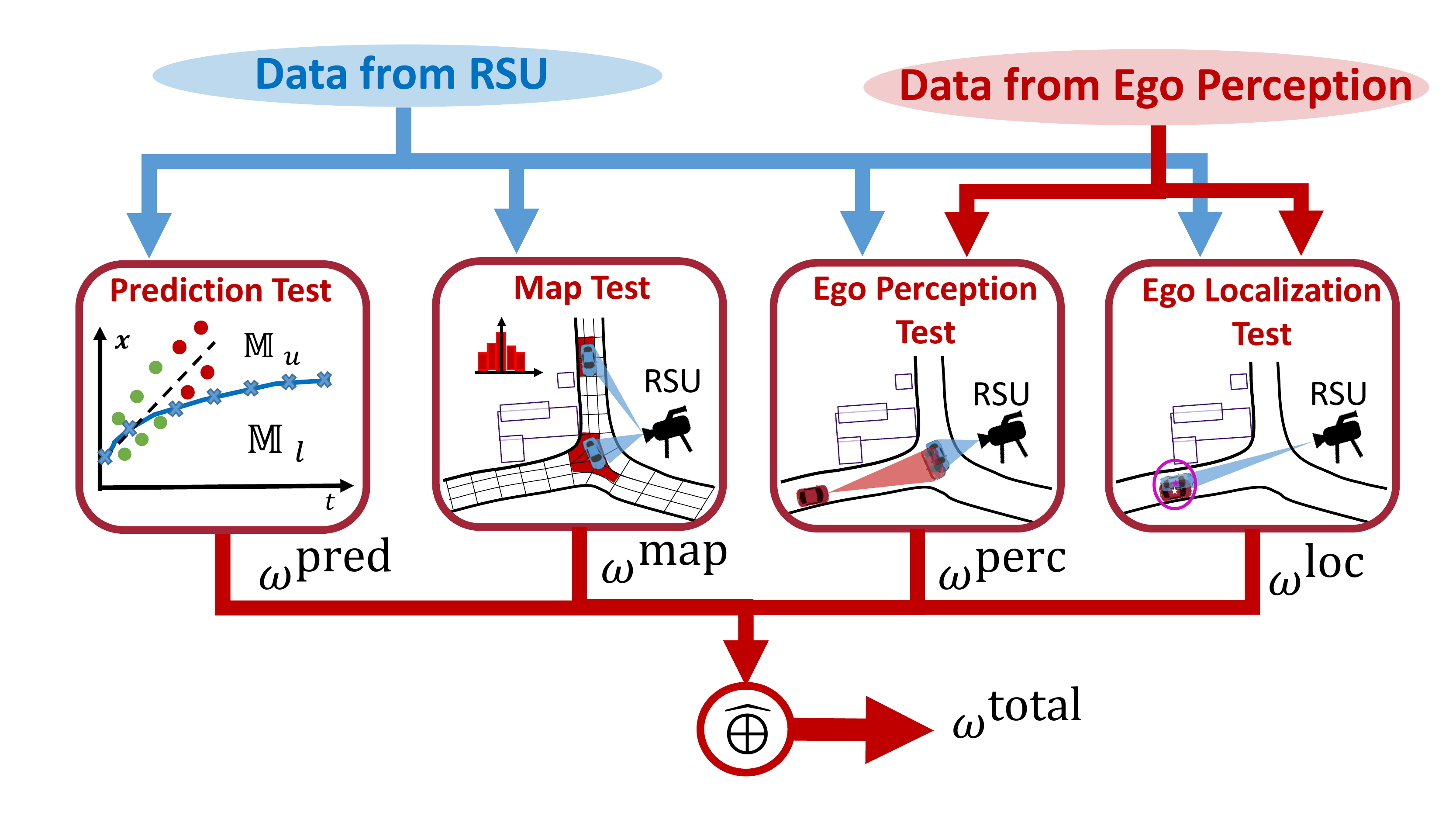}
	\caption{System overview}
	\label{fig:Overview}
\end{figure}

In this work, a Subjective-Logic-based (SL-based) online estimation mechanism is presented, that holistically estimates the reliability of cooperative information reported by other agents, such as the RSU 
in the example mentioned above.
Through the use of SL calculus, IVs are not only able to calculate the SOTIF-related measures, but also determine their corresponding certainty in terms of statistics.
Testing for different aspects of functionality results in multiple information sources within the IV, termed as opinions \cite{Joesang2016}. By SL-based fusion of these opinions, we are able to monitor and incorporate multiple areas of functional errors.
In this work, we propose the use of four opinions, each providing evidence for the correctness of the overall system. The four tests are 1) the prediction test, in which the RSU's past behavior prediction of individual objects within its field of view (FOV) is compared to the current situation, 2) the map test, which checks the reported objects for consistency with the IV's digital map, 3) the ego perception test, which compares detections from the IV's ego perception with the reported object list, and 4) the ego localization test, which compares the estimated ego position with that of the corresponding reported object along with its uncertainty. The system overview is sketched in Fig. \ref{fig:Overview}.

In distributed systems theory, integrity monitoring and reliability estimation is generally termed \emph{Fault Detection and Isolation} (FDI) \cite{Zidan2017}. However, while being a long investigated topic, most developed mechanisms still suffer from exhaustive resource consumption and are centralized by nature, i.e. require an overall monitoring entity and are thus unable to scale up to large installations \cite{Zidan2017},\cite{Dorri2018}. Furthermore, isolation of (identified) faulty nodes is an open issue \cite{Dorri2018}. The scheme described within this work shows an online-capable and thus low resource consuming, decentralized approach, that is able to detect faulty behavior before the fault takes effect.

In the automotive domain, integrity monitoring for IVs is used to monitor digital map reliability using a data-driven classifier approach \cite{Hartmann2014} and to estimate and assure lane information quality \cite{Nguyen2016}, including probabilistic fusion from multiple sources using the Dempster-Shafer (DS) theory of evidence \cite{Nguyen2017},\cite{Nguyen2018}. While being effective in practice, for their specific use-case, these approaches are tailored towards use of data that is generated from on-board sensors within an IV. We, on the other hand, propose a solution that uses information from the outside and additionally is simple in terms of computational complexity. Lastly, our approach relies on the theory of SL \cite{Joesang2016}, thus does not suffer from the drawback of inconsistent system behavior in some cases due to weaknesses in the underlying DS-based fusion \cite{Nguyen2017},\cite{Nguyen2018}.
In comparison to \cite{Obst2014}, our system takes into account multiple information sources and yields continuous reliabilities as well as their estimation uncertainties rather than a binary classification as part of the fusion process. 

Recent works on misbehavior detection in vehicular ad-hoc networks (VANETS) proposed the use of neural networks \cite{Ghaleb2017} or SL \cite{Dietzel2014}. While also being designed for large scale and distributed systems, both approaches focus on the security aspects of VANETS. While we acknowledge the importance of security for intra-vehicle communication, we focus on the safety of vehicular communication and the corresponding SOTIF challenges, i.e. assuring functionality safety of a cooperatively perceived environmental model. 

In light of previous research, we propose a novel framework for online reliability estimation of cooperative information received from RSUs using the theory of Subjective Logic. The contribution of this paper is twofold. First, we show how SL can be used as a general framework for IVs to assess and fuse individual, orthogonal information cues into one holistic overall reliability estimate and guarantee SOTIF protection levels. Second, we propose four exemplary tests in the context of IVs receiving cooperative information from an RSU. These tests are designed to enable a SOTIF-aware functionality for IVs and are tested in a real-world experiment, which shows that the proposed framework delivers reliable and good results. 


\section{SUBJECTIVE LOGIC BASICS} \label{SL_basics}

Subjective Logic is an extension of classical probability theory that explicitly models statistical uncertainty. It unifies several evidence-based extensions of classical probability theory, such as Dempster-Shafer theory, and is linked to classical probability theory by a bijective mapping to Dirichlet distributions \cite{Joesang2016}.
One of the key features of SL is a set of fusion operators that make it possible to combine different pieces of evidence, termed opinions. The Aleatory Cumulative Belief Fusion operator is used to combine independent pieces of evidence to reduce the statistical uncertainty. In turn, the Aleatory Belief Fusion operator is used to average over statistical dependent pieces of evidence. Finally, the Uncertainty Weighted Average Belief Fusion operator does a weighted averaging between two opinions, while the result tends towards the more certain opinion.
In this section, we briefly summarize the SL basics used in this paper. The definitions and theorems are taken from \cite{Joesang2016}, where further details can be found.

\textbf{Definition 1 (Subjective Logic opinion):} Let $\mathbb{X}$ be a domain and $\text{card}\{\mathbb{X}\} \geq 2$, where $\text{card}\{ \, \cdot \, \}$ is the cardinality. Let $X$ further be a random variable in $\mathbb{X}$. A SL opinion (opinion in short) is an ordered triple $\omega_x = (\boldsymbol{b}_x,u_x,\boldsymbol{a}_x)$ with
\begin{subequations}
	\begin{align}
	\boldsymbol{a}_x : \mathbb{X} \mapsto [0,1], \qquad 1 &= \sum\limits_{x \in \mathbb{X}} \boldsymbol{a}_x \, ,\\
	\boldsymbol{b}_x : \mathbb{X} \mapsto [0,1], \qquad 1 &= u_x + \sum\limits_{x \in \mathbb{X}} \boldsymbol{b}_x \, .
	\end{align}
\end{subequations}
Hereby, $\boldsymbol{b}_x$ is the belief mass distribution over $\mathbb{X}$, $u_x$ is the uncertainty mass representing a lack of evidence, and $\boldsymbol{a}_x$ is the base rate distribution over $\mathbb{X}$ representing the prior. 

\textbf{Definition 2 (Dirichlet Distribution):} Let $\mathbb{X}$ be a domain of $W$ mutually disjoint values, $\boldsymbol{r}_x \in \mathbb{N}$ be the evidence for outcome $x \in \mathbb{X}$, $\boldsymbol{a}_x$ a prior distribution over $\mathbb{X}$, and $\boldsymbol{p}_x$ the probability distribution of $x$ over $\mathbb{X}$. Then, the probability density function (PDF)
\begin{equation}
\text{Dir}(\boldsymbol{p}_x,\boldsymbol{r}_x,\boldsymbol{a}_x) = \frac{\Gamma \left( \sum \limits_{x \in \mathbb{X}} (\boldsymbol{r}_x + \boldsymbol{a}_x W) \right)}{\prod \limits_{x \in \mathbb{X}} \Gamma ( \boldsymbol{r}_x + \boldsymbol{a}_x W ) } \prod \limits_{x \in \mathbb{X} } \boldsymbol{p}_x ^{\boldsymbol{r}_x + \boldsymbol{a}_x W - 1} \, ,
\label{eq:Dirichlet}
\end{equation} 
where $\boldsymbol{r}_x + \boldsymbol{a}_x W \geq 0$ and $\boldsymbol{p}_x > 0$ for $\boldsymbol{r}_x + \boldsymbol{a}_x W < 1$, is called Dirichlet PDF. A Dirichlet PDF with $W=2$ is called $\beta$-distribution. In (\ref{eq:Dirichlet}), $\Gamma( \, \cdot \, )$ is the Gamma function \cite{Bronshtein2007}.

\textit{Remark:} Dirichlet PDFs, in essence, are probability distributions over discrete probability distributions, stating the probability that according to the available evidence, the assumed discrete probability distribution is correct. Thus, Dirichlet PDFs and $\beta$-PDFs in particular are used within this work to visualize SL opinions as well as reliabilities. The higher the value of a $\beta$-PDF at a given probability, the more evidence supports the probability.

\textbf{Definition 3 (Aleatory Cumulative Belief Fusion):} 
Let $\omega_x^A$ and $\omega_x^B$ be sources A's and B's respective opinions over the same variable $X$ on domain $\mathbb{X}$. Then, the operator $\oplus$ in
\begin{equation}
\omega_x^A \oplus \omega_x^B = \left \{  \begin{array}{ll}
\boldsymbol{b}_x &= \frac{ \boldsymbol{b}_x^{A} u_x^{B} + \boldsymbol{b}_x^{B} u_x^{A} }{ u_x^{A} + u_x^{B} - u_x^{A} u_x^{B} } \\
u_x &= \frac{ u_x^{A} u_x^{B} }{ u_x^{A} + u_x^{B} - u_x^{A} u_x^{B} }\\
\boldsymbol{a}_x &= \frac{ \boldsymbol{a}_x^{A} u_x^{B} + \boldsymbol{a}_x^{B} u_x^{A} - (\boldsymbol{a}_x^{A} + \boldsymbol{a}_x^{B}) u_x^{A} u_x^{B} }{ u_x^{A} + u_x^{B} - 2 u_x^{A} u_x^{B} } 	
\end{array}\right . \, ,
\end{equation}
where $0 < u_x^A < 1$ and $0 < u_x^B < 1$, is called Aleatory Cumulative Belief Fusion.

\textbf{Definition 4 (Aleatory Average Belief Fusion):} 
Let $\omega_x^A$ and $\omega_x^B$ be sources A's and B's respective opinions over the same variable $X$ on domain $\mathbb{X}$. Then, the operator $\underline{\oplus}$ in
\begin{equation}
\omega_x^A \, \underline{\oplus} \, \omega_x^B = \left \{  \begin{array}{ll}
\boldsymbol{b}_x &= \frac{ \boldsymbol{b}_x^{A} u_x^{B} + \boldsymbol{b}_x^{B} u_x^{A} }{  u_x^{A} \, + \, u_x^{B} }  \\
u_x &= \frac{2 u_x^{A} u_x^{B} }{ u_x^{A} \, + \, u_x^{B} }  \\
\boldsymbol{a}_x &= \frac{ \boldsymbol{a}_x^{A} \, + \, \boldsymbol{a}_x^{B} }{ 2 } 	
\end{array}\right . \, ,
\end{equation}
where $u_x^A \neq 0$ and $u_x^B \neq 0$, is called Aleatory Average Belief Fusion. When multiple opinions are fused, we use the shorthand 
\begin{equation}
\overset{N}{\underset{i=1} {\underline{\bigoplus}} } \, \omega_x^{A_i} = \omega_x^{A_1} \, \underline{\oplus} \, \omega_x^{A_2} \underline{\oplus} \dots \underline{\oplus}  \omega_x^{A_N} \, .
\end{equation}

\textbf{Definition 5 (Uncertainty Weighted Average Belief Fusion):} 
Let $\omega_x^A$ and $\omega_x^B$ be sources A's and B's respective opinions over the same variable $X$ on domain $\mathbb{X}$. Then, the operator $\hat{\oplus}$ in
\begin{equation}
	\omega_x^{A} \; \hat{\oplus} \; \omega_x^{B} = \left \{ \begin{array}{ll}
	\boldsymbol{b}_x^{A \diamond B} = \frac{ \boldsymbol{b}_x^{A} (1-u_x^{A}) u_x^{B} + \boldsymbol{b}_x^{B} (1-u_x^{B}) u_x^{A} }{ u_x^{A} + u_x^{B} - 2 u_x^{A} u_x^{B} }  \\
	u_x^{A \diamond B} = \frac{ (2 - u_x^{A} - u_x^{B} u_x^{A} u_x^{B} ) }{ u_x^{A} + u_x^{B} - 2 u_x^{A} u_x^{B} }  \\
	\boldsymbol{a}_x^{A \diamond B} = \frac{ \boldsymbol{a}_x^{A}(1-u_x^{A}) + \boldsymbol{a}_x^{B}(1-u_x^{B})  }{2 - u_x^{A} - u_x^{B} }
	\end{array} \right .
\end{equation}
with $(u_x^{A} \neq 0 \vee u_x^{B} \neq 0 ) \wedge ( u_x^{A} \neq 1 \vee u_x^{B} \neq 1 )$, is called Uncertainty Weighted Average Belief Fusion.

\textbf{Theorem 1 (Equivalent mapping):} Let $\omega_x = (\boldsymbol{b}_x,u_x,\boldsymbol{a}_x)$ be an opinion and $\text{Dir}(\boldsymbol{p}_x,\boldsymbol{r}_x,\boldsymbol{a}_x)$ a Dirichlet distribution over the same $x \in \mathbb{X}$. Then, the 
mapping
\begin{equation}
\label{eq:Mapping}
\!\!\! \left . \begin{array}{l}
\boldsymbol{b}_x = \frac{\boldsymbol{r}_x}{W + \sum \limits_{x_i \in \mathbb{X}} \boldsymbol{r}_x} \\
u_x = \frac{W}{W + \sum \limits_{x_i \in \mathbb{X}} \boldsymbol{r}_x} 
\end{array} \right \} \!\!\! \iff \!\!\!
\left \{ \begin{array}{ll}
\boldsymbol{r}_x &\!\!\!= \frac{ W \boldsymbol{b}_x}{u_x} \\
1 &\!\!\!= u_x + \sum \limits_{x_i \in \mathbb{X}} \boldsymbol{b}_x(x_i)
\end{array} \right . 
\end{equation}
transforms the Dirichlet PDF into the opinion and vice versa.
\textit{Proof:} See \cite{Joesang2016}.


\section{RELIABILITY ESTIMATION MECHANISM}

In this section, the reliability estimation mechanism is described in detail. First, the four single tests are described. Then, the fusion process of all test results is explained.

\subsection{Consistency Check of the Prediction}
The basic idea of the consistency check of the RSU's behavior prediction is to accumulate predictions as well as the actual measurements and then compare whether or not the former predictions fit to the latter measurements. 
Therefore, upper and lower bounds of each prediction point are calculated. Mathematically, this forms a set of tuples $\{ (\boldsymbol{x}_{p,l}, \, \boldsymbol{x}_{p,u})_i \}$, where $\boldsymbol{x}_{p,l}$ is the lower bound of a prediction point and $\boldsymbol{x}_{p,u}$ is the upper bound, respectively. 
In the second step, the set of actual measurements $\{  \boldsymbol{x}_{m,j} \}$ is linearly interpolated resulting in the polygon line $\mathcal{PL}(\{  \boldsymbol{x}_{m,j} \})$. The polygon line then splits the measurement space $\mathbb{M}$ into an upper and a lower subspace $\mathbb{M}_u(\{  \boldsymbol{x}_{m,j} \})$ and $\mathbb{M}_l(\{  \boldsymbol{x}_{m,j} \})$, respectively. The set of prediction tuples $\{ (\boldsymbol{x}_{p,l}, \, \boldsymbol{x}_{p,u})_i \}$ then is partitioned into a correct and incorrect subset:
\begin{subequations}
	\begin{align}
	\mathcal{S}_{\text{correct}} &= \{ (\boldsymbol{x}_{p,l}, \, \boldsymbol{x}_{p,u})_i | \boldsymbol{x}_{p,l} \in \mathbb{M}_l \wedge \boldsymbol{x}_{p,u} \in \mathbb{M}_u \} \, , \\
	\mathcal{S}_{\text{incorrect}} &= \{ (\boldsymbol{x}_{p,l}, \, \boldsymbol{x}_{p,u})_i \} \setminus \mathcal{S}_{\text{correct}} \, .
	\end{align}
\end{subequations}
With the evidence vector
\begin{equation}
\boldsymbol{r}_x = \begin{bmatrix}
\text{card}\{ \mathcal{S}_{\text{correct}} \} \\ 
\text{card}\{ \mathcal{S}_{\text{incorrect}} \}
\end{bmatrix}
\end{equation}
and the equivalent mapping (\ref{eq:Mapping}), the partitioned subsets are mapped to the opinion $\tilde{\omega}^{\text{pred}}$. Finally, a probability sensitive trust discounting
\begin{equation}
\label{eq:TrustDiscountingPrediction}	
\omega^{\text{pred}} = \left \{ \begin{array}{ll}
\boldsymbol{b}^{\text{pred}} &= \boldsymbol{p}_{\text{indep}} \; \tilde{\boldsymbol{b}}^{\text{pred}} \\
u^{\text{pred}} &= 1 - \boldsymbol{p}_{\text{indep}} \sum \limits_{x \in \mathcal{R}(\mathbb{X})}  \tilde{\boldsymbol{b}}^{\text{pred}} \\
\boldsymbol{a}^{\text{pred}} &= \boldsymbol{a}^{\text{pred}} 
\end{array}\right . \\	
\end{equation}
of the test opinion, accounting for the statistical dependence of the prediction samples, is performed. Herby, $\boldsymbol{p}_{\text{indep}}$ can be interpreted as the probability vector that the current piece of evidence is statistically independent of the evidence already accounted for and thus truly brings new information into the fusion result. 
Simply spoken, a tuple is classified as correct, if the polygon line intersects the corresponding interval. Otherwise, it is classified as incorrect.
Note that the trust discounting step basically has the same effect as applying the partially dependent source cumulative belief fusion operator from SL described in \cite{Josang2006}.

As we focus on reliability analysis rather than object prediction in this work, for simplicity, a Kalman filter with a constant velocity model is used to predict the objects' future positions along their lanes. This is then used as behavior prediction of the RSU. Thus, the upper and lower bounds can simply be chosen as the $\pm \sigma$  bounds of the individual prediction points.
Note that more sophisticated methods for behavior prediction, e.g. \cite{Wiest2012}, exist and can be integrated as well. The upper and lower bounds then have to be chosen accordingly.

\subsection{Consistency Check Between Map and Received Object Lists}
The idea of this test is to detect inconsistency between the digital map of the IV and the object list $\mathcal{O}_{\text{RSU}}$ reported by the RSU. Therefore, the typical positions of the object centers, according to the RSU perception, are determined during commissioning, 
hence guaranteeing the correct functionality of the RSU.
This is done using several 1D histograms along the lanes spread over the RSU's FOV, where the 1D histograms only resolve the lateral position of the object centers and are simply stacked along the lanes (see Fig. \ref{fig:FOV}). 
The individual bin length $l_{bin}$ is chosen to compromise between fine-grained approximation of the corresponding continuous density and a reasonable value of object probes that accumulates within each bin. Note that we do not expect the exact number of bins to have a strong influence on the overall result because of the Aleatoric Average Belief Fusion performed over all individual bins. The reference histograms then are mapped to reference opinions $\omega_x^{\text{ref},i}$ using the equivalence mapping (\ref{eq:Mapping}) and pooled according to
\begin{equation}
	\omega_x^{\text{ref}} = {\underset{\forall i} {\underline{\bigoplus}} } \, \omega_x^{\text{ref},i} \, .
\end{equation}
Thus, only one average opinion $\omega_x^{\text{ref}}$ per lane is computed. These $\omega_x^{\text{ref}}$ are stored in the digital map of the vehicle.

When the IV approaches the FOV of the RSU, it starts to create a set of histograms using the same bins as for the creation of the reference opinion. The sample opinion
\begin{equation}
\omega_x^{s} = {\underset{\forall i} {\underline{\bigoplus}} } \, \omega_x^{s,i}
\end{equation}
is then determined online, the same way as $\omega_x^{\text{ref}}$ but with only the online data available from the reported object lists $\mathcal{O}_{\text{RSU},k}$. If the RSU data is consistent with the map, the distance between the opinions $\omega_x^{s}$ and $\omega_x^{\text{ref}}$ should be small, i.e. below a threshold $\theta_{\textit{DC}}$. A measure for the distance between opinions is the \textit{degree of conflict}\cite{Joesang2016}
\begin{equation}
\textit{DC}_x^{s,\text{ref}} = \frac{1}{2} \sum \limits_{\forall x \in \mathbb{X}} | \boldsymbol{p}_{x}^{s} - \boldsymbol{p}_{x}^{\text{ref}} | (1 - u_x^{\text{ref}}) (1 - u_x^{s})\, ,
\end{equation}
where $\boldsymbol{p}_{x}^{s} = \boldsymbol{b}_{x}^{s} + u_{x}^{s} \boldsymbol{a}_{x}^{s}$ and $\boldsymbol{p}_{x}^{\text{ref}} = \boldsymbol{b}_{x}^{\text{ref}} + u_{x}^{\text{ref}} \boldsymbol{a}_{x}^{\text{ref}}$, respectively, are the maximum-likelihood mappings of the opinions $\omega_x^{s}$ and $\omega_x^{\text{ref}}$ to probabilities \cite{Joesang2016}.

If $\textit{DC}_x^{s,\text{ref}} < \theta_{\textit{DC}}$, the incoming RSU data is considered to match the vehicle's map and hence the estimated reliability within the map test opinion $\omega^{\text{map}}$ is increased according to
\begin{subequations}
	\label{eq:TrustIncreaseMap}
	\begin{align}
		\omega^{\text{evid}}(k) &= \left(\begin{bmatrix}
		\frac{1}{W+1} \\
		0
		\end{bmatrix}, \, \frac{W}{W + 1}, \, \begin{bmatrix}
		0.5 \\
		0.5
		\end{bmatrix} \right) \, , \\	
		\tilde{\omega}^{\text{evid}} &= \left \{ \begin{array}{ll}
		\boldsymbol{b}^{\text{evid}} &= \boldsymbol{p}_{\text{dis}} \boldsymbol{b}^{\text{evid}} \\
		u^{\text{evid}} &= 1 - \boldsymbol{p}_{\text{dis}} \sum \limits_{x \in \mathcal{R}(\mathbb{X})}  \boldsymbol{b}^{\text{evid}} \\
		\boldsymbol{a}^{\text{evid}} &= \boldsymbol{a}^{\text{evid}} 
		\end{array}\right . \, ,\\	
		\omega^{\text{map}}(k+1) &=\omega^{\text{map}}(k) \oplus \tilde{\omega}^{\text{evid}}(k)  \, ,
	\end{align}
\end{subequations}
where $\omega^{\text{evid}}$ is the opinion generated from the new evidence, and $\tilde{\omega}^{\text{evid}}$ is the probability sensitive trust discounted opinion \cite{Joesang2016} about the new evidence. $\boldsymbol{p}_{\text{dis}}$ can be interpreted as the probability that a single measurement is meaningful. Here, $\boldsymbol{p}_{\text{dis}} = [0.1, \, 0.1]^{\top}$ is used.
Otherwise, the trust into the RSU is revised according to the probability sensitive trust revision \cite{Joesang2016}:
\begin{equation}
	\omega^{\text{map}}(k+1) = \left \{ \begin{array}{l}
	\tilde{\boldsymbol{b}}(k+1) = (1- DC_x^{s,\text{ref}})\boldsymbol{b}(k) \\
	\tilde{u}(k+1) = (1- \textit{DC}_x^{s,\text{ref}})u(k) \\
	\tilde{\boldsymbol{a}}(k+1) = \boldsymbol{a}(k)
	\end{array} \right . \, .
\end{equation}

\subsection{Comparison Between Ego Perception and Received Object Lists}
The RSU's main purpose is to extend the environmental model of the IV into areas that are occluded for the IV's ego perception. Therefore, it can be expected that the mutual FOV of the IV and the RSU is small and mostly in corner regions, where the IV's ego perception reliability is reduced. Furthermore, objects may move quickly through the narrow mutual FOV. Thus, it can be expected that the IV's ego perception will only receive very view measurements to confirm the existence of an object reported by the RSU. 
As a first step, we perform nearest neighbor association with the Euclidean distance between object $o_i^{RSU}$ received from the RSU and the objects $o_i^{ego}$ detected with the IV's on-board perception FOV. For gating, a simple thresholding with a fixed threshold $d_{max}$ is used before association. Note that more sophisticated mechanisms for both gating and data association are available \cite{Reuter2014} and could be integrated as well. We, however, want to focus on the overall idea of SL-based fusion and reliability estimation in this work and thus use this simple, yet effective technique for gating and association.
If a matching object is found, the trust into the RSU is increased, else a missing detection is assumed and the trust into the RSU is strongly reduced. 
Mathematically, this mechanism can be described as
\begin{subequations}
	\label{eq:EgoPerception}
	\begin{align}
	\omega^{\text{conf}}(k) &= \left(\begin{bmatrix}
	\frac{1}{W+1} \\
	0
	\end{bmatrix}, \, \frac{W}{W + 1}, \, \begin{bmatrix}
	0.5 \\
	0.5
	\end{bmatrix} \right) \, , \\	
	\omega^{\text{mis}}(k) &= \left(\begin{bmatrix}
	0 \\
	\frac{w_{\text{mis}}}{W+w_{\text{mis}}}
	\end{bmatrix}, \, \frac{W}{W + w_{\text{mis}}}, \, \begin{bmatrix}
	0.5 \\
	0.5
	\end{bmatrix} \right) \, , \\
	\omega^{\text{input}}(k) &= \left \{ \begin{array}{ll}
	\omega^{\text{conf}}(k) \, &\text{if } \exists o_i | d_E(o_{\text{ego}},o_i) < d_{\text{max}}  \, , \\
	\omega^{\text{mis}}(k) \, &\text{otherwise, }
	\end{array} \right . \\
	\omega^{\text{perc}}(k+1) &=\omega^{\text{perc}}(k) \oplus \omega^{\text{input}}(k)  \, .
	\end{align}
\end{subequations}
In (\ref{eq:EgoPerception}), $\omega^{\text{conf}}$ describes the opinion resulting from an object within the ego perception that confirms an object $o_i$ from the RSU's object list $\mathcal{O}_{\text{RSU}}$, while $\omega^{\text{mis}}$ is the opinion resulting from a missing detection, i.e. when the ego perception detects an object that has no match in $\mathcal{O}_{\text{RSU}}$. 
To account for the fact that missing detection might have severe implications, they are up-weighted and compared to positive tests with the parameter $w_{\text{mis}}$, which ensures that missing objects decrease confidence in the RSU significantly.

The new input opinion $\omega^{\text{input}}$ then is fused with the last perception test opinion $\omega^{\text{perc}}(k)$ resulting in the new perception test opinion $\omega^{\text{perc}}(k+1)$. This time-recursive structure incorporates the fact that both the ego and the RSU perception are expected to have recursive estimators to create their individual object lists.

\subsection{Comparison Between Ego Localization and Received Object Lists}
Even though the comparison between ego perception and received object list can detect 
the occurrence of missed detections in the RSU, 
it does not make a statement on the measurement uncertainties reported by the RSU.
This is due to the fact that the measured objects are extended by nature. Thus, the IV may perceive only a part of an object, while the RSU perceives the whole object resulting in differing object center estimates. A check on the uncertainties thus would not always be meaningful for the ego perception test.
Hence, an additional test is needed. By comparing the object $o_{\text{RSU,ego}} \in \mathcal{O}_{\text{RSU}}$ corresponding to the IV with the ego localization, a very precise measurement is available to evaluate the uncertainties reported by the RSU. Furthermore, the ego localization test is always possible, even if no other road user is approaching the intersection. 

The reported uncertainties are checked by comparing the Euclidean distance between ego localization and object center of the corresponding RSU object $d_E(x_{\text{ego}},x_{o_{\text{ego}}})$ with the total uncertainty consisting of the ego localization uncertainty and the reported uncertainty $\sigma_{\text{total}} = \sigma_{o_{\text{ego}}} + \sigma_{\text{ego}}$. If $d_E(x_{\text{ego}},x_{o_{\text{ego}}}) < 3 \cdot \sigma_{\text{total}}$, the opinions are considered consistent and the trust into the RSU is increased. 
Otherwise, under the assumption that the error is Gaussian, the probability of the data being consistent is less than $1\%$. 
Thus, in this case, the trust into the RSU is reduced. If no matching object $o_{\text{ego}}$ can be found within $d_{\text{max}}$, a missing detection in the RSU's perception is detected and the incidence is passed to the ego perception test. Mathematically, the ego localization test can be formulated as
\begin{subequations}
	\label{eq:EgoLocalization}
	\begin{align}
	\omega^{\text{conf}}(k) &= \left(\begin{bmatrix}
	\frac{1}{W+1} \\
	0
	\end{bmatrix}, \, \frac{W}{W + 1}, \, \begin{bmatrix}
	0.5 \\
	0.5
	\end{bmatrix} \right) \, , \\	
	\omega^{\text{under}}(k) &= \left(\begin{bmatrix}
	0 \\
	\frac{w_{\text{under}}}{W+w_{\text{under}}}
	\end{bmatrix}, \, \frac{W}{W + w_{\text{under}}}, \, \begin{bmatrix}
	0.5 \\
	0.5
	\end{bmatrix} \right) \, , \\
	\omega^{\text{input}}(k) &= \left \{ \begin{array}{ll}
	\omega^{\text{conf}}(k) \, &\!\!\! \text{if }d_E(x_{\text{ego}},x_{o_{\text{ego}}}) < 3 \cdot \sigma_{\text{total}} \, , \\
	\omega^{\text{under}}(k) \, &\!\!\! \text{otherwise, }
	\end{array} \right . \\
	\omega^{\text{loc}}(k+1) &=\omega^{\text{loc}}(k) \oplus \omega^{\text{input}}(k)  \, ,
	\end{align}
\end{subequations} 
where $\omega^{\text{conf}}$ is the opinion confirming the RSU, $\omega^{\text{under}}$ is the opinion resulting from an underestimated uncertainty and $\omega^{\text{input}}$ is the new input opinion that the ego localization test opinion $\omega^{\text{loc}}$ is updated with.

\subsection{Fusion of the Test Results}
So far, four tests have been proposed, each of which focuses on a different property of valid RSU data. While the map test and the prediction test completely rely on the data reported by the RSU, the ego perception test and the ego localization test include the IV's perception as an additional source of information. Thus, given the same amount of evidence, the tests including the IV's perception are far more important for the overall estimation of the RSU's reliability. In turn, as the ego perception test as well as the ego localization test need the IV to be already very close to the intersection, these tests will usually operate on a far smaller amount of evidence than the tests independent of the IV's perception. Both effects are accounted for by combining the uncertainty weighted average belief fusion operator with a simple importance weighted average belief fusion. Mathematically, this can be expressed as
\begin{subequations}
	\label{eq:TestFusion}
	\begin{align}
		\omega^{\text{it}} &= \omega^{\text{pred}} \; \hat{\oplus} \; \omega^{\text{map}} \overset{\text{(\ref{eq:Mapping})}}{\implies} 
		\boldsymbol{r}^{\text{it}} \, , \\
		\omega^{\text{ept}} &= \omega^{\text{perc}} \; \hat{\oplus} \; \omega^{\text{loc}} \overset{\text{(\ref{eq:Mapping})}}{\implies} 
		\boldsymbol{r}^{\text{ept}} \, , \\	
		\boldsymbol{r}^{\text{total}} &= \frac{w_{\text{ept}} \; \boldsymbol{r}^{\text{it}} + w_{\text{it}} \; \boldsymbol{r}^{\text{it}} }{ w_{\text{ept}} + w_{\text{it}} } \, , \\
		\omega^{\text{total}} &\overset{ \text{(\ref{eq:Mapping})} }{=} \left \{ \begin{array}{ll}
		\boldsymbol{b}^{\text{total}} &\!\!\!= \frac{\boldsymbol{r}^{\text{total}} }{ W + \sum \limits_{x \in \mathbb{X}} \boldsymbol{r}^{\text{total}} } \\
		u^{\text{total}} &\!\!\!= \frac{W}{ W + \sum \limits_{x \in \mathbb{X}} \boldsymbol{r}^{\text{total}} } \\
		\boldsymbol{a}^{\text{total}} &\!\!\!= \frac{w_{\text{it}} \boldsymbol{a}^{\text{it}} \, + \, w_{\text{ept}} \boldsymbol{a}^{\text{ept}} }{ w_{\text{it}} \, + \, w_{\text{ept}} } 
		\end{array} \right . \, .
	\end{align}
\end{subequations}
In (\ref{eq:TestFusion}), $\omega^{\text{it}}$ describes the fused opinion on the ego perception independent tests, while $\omega^{\text{ept}}$ describes the fusion of the ego-perception-based tests. 
The opinions $\omega^{\text{it}}$ and $\omega^{\text{ept}}$ are then mapped to the Dirichlet space using the mapping (\ref{eq:Mapping}), weighted there with $w_{\text{it}}$ and $w_{\text{ept}}$, respectively, and mapped back to the overall fusion result  $\omega^{\text{total}}$. Thus, (\ref{eq:TestFusion}) defines a new SL fusion operator that does a belief and importance weighted average belief fusion.

\section{EXPERIMENTAL SETUP}

\begin{figure}[bt]
	\centering
	\includegraphics[width=0.22\textwidth]{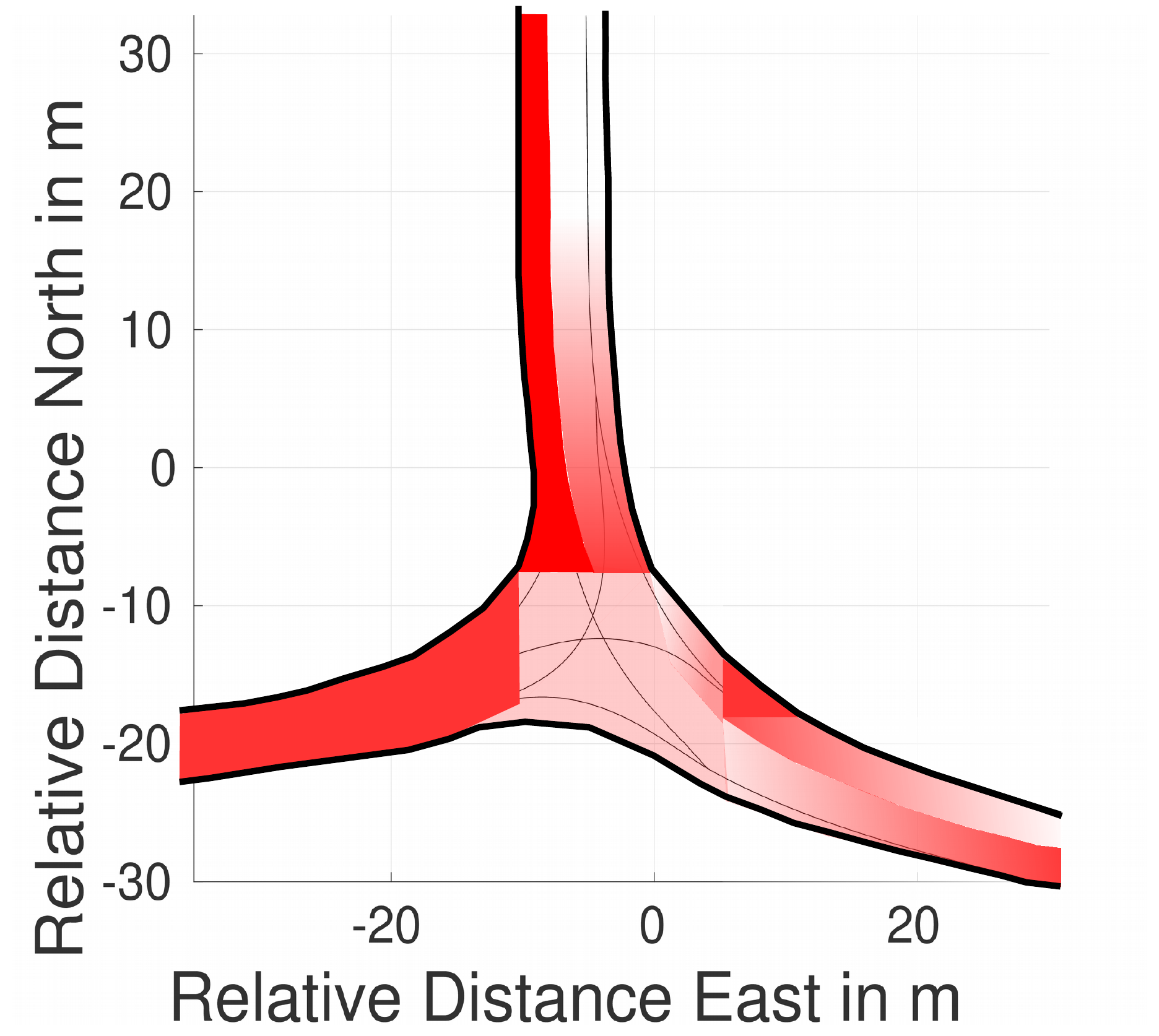}
	\includegraphics[width=0.25\textwidth]{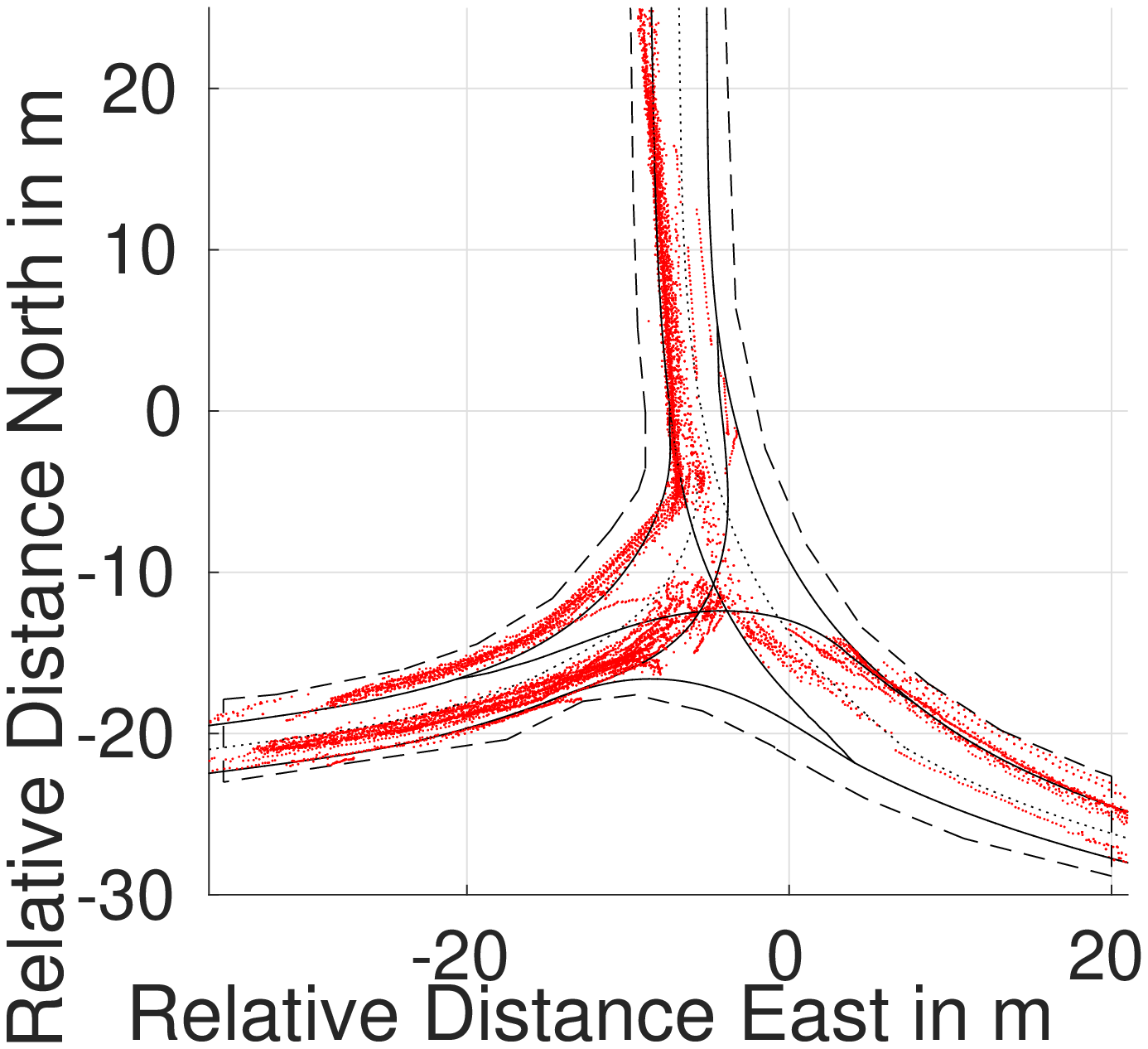}
	\caption{Field of View of the RSU. Left, the RSU's reliability is sketched, where darker means more reliable. On the right a map of measured object centers is depicted. The object centers are marked red. As can be seen, the sensor coverage in the central area of the intersection is reduced.}
	\label{fig:FOV}
\end{figure}

The experiment is set up at a mid-size city crossing, with occlusion due to close-to-road buildings. Besides others, four SICK LD-MRS 8 layer laserscanners are mounted on poles in $7 \, \text{m}$ height, observing the intersection. Their data is fused and the resulting object list is then reported to the IV by the RSU. Fig. \ref{fig:FOV} sketches the FOV covered by the LiDAR subsystem of the RSU. As can be seen, the laserscanners show a blind spot where sensor reliability is reduced\footnote{The blind spot is observed by the other sensors, not used here.}. Thus, real examples of missing detections are expected to be found within the recorded data. The IV approaches this intersection, while the RSU sends object lists of its perception.

For the IV, an automated vehicle as described in \cite{Kunz2015} is used. For the ego perception, the front IBEO 4 layer laserscanner is used. The IV is localized using a high precision real time kinematic (RTK) system.


\section{EVALUATION}
 For evaluation, $14$ sequences where recorded on two different days, where the IV approaches the intersection on different entrance lanes. 
The sequences have been manually labeled with respect to malfunctions of the RSU such as missing detections. In $7$ sequences the RSU has been manually classified as correctly behaving, while in the other $7$, the RSU shows misbehavior.
The map test could not, however, be evaluated without modifying the recorded data. In order to evaluate the map test, thus, the original data reported by the RSU are shifted in east direction. This corresponds to a differently chosen reference point or a calibration error due to a GPS offset during recalibration. 
 
\subsection{Consistency Check Evaluation of the Prediction} 
\begin{figure}
 	\centering
 	\includegraphics[width=0.4\textwidth]{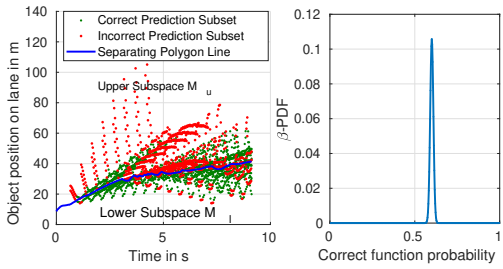}
 	\caption{Prediction test: The left picture shows the partitioning of the measurement space as well as the classification into correctly predicted tuples and incorrect predicted tuples. In the right picture, the resulting $\beta$-PDF is depicted. As a lot of tuples are incorrectly predicted, the $\beta$-PDF is shifted to about $60\%$ correct function probability indicating a reduced reliability.}
 	\label{fig:Prediction}
 \end{figure}

 Figure \ref{fig:Prediction} exemplary visualizes the result of the prediction test. In the given example, the observed vehicle decelerates. Hence, some of the prediction tuples $\{ (\boldsymbol{x}_{p,l}, \, \boldsymbol{x}_{p,u})_i \}$ are classified as incorrect. This leads to a reduction of the estimated reliability. Thus, it can be seen that the proposed mechanism behaves as expected.
 
\subsection{Evaluation of the Consistency Check Between Map and Received Object Lists}

\begin{figure}[htb]
	\centering
	\includegraphics[width=0.45\textwidth]{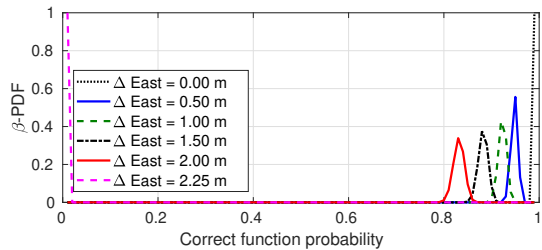}
	\caption{Map test opinion after shifting the RSU objects in east direction. As can be seen, depending on the map-shift amount, the reliability estimate drops from higher to lower values, consistent with the expectations in the experimental setup.}
	\label{fig:MapEast}
\end{figure}

To evaluate the consistency test between the map and the received object lists, the objects from the RSU are shifted by a constant offset to simulate an RSU calibration or mapping error. 
As the object centers are highly concentrated within a small band in the northern branch of the intersection (see Fig. \ref{fig:FOV}), the probability gradient along east direction is big. 
Figure \ref{fig:MapEast} shows the effect of a shift in east direction on the estimated reliability. It can be seen that the test reacts quite sensitively to a shift in east direction due to the big probability gradient.
If the shift is big enough, the RSU reports objects at positions that are physically unfeasible, i.e. objects moving over houses. Thus, the estimated reliability then is reduced to zero all of a sudden (see $\Delta \text{East} = 2.25 \,\text{m}$). 

\subsection{Evaluation of the Comparison between Ego Perception and Received Object Lists}  

\begin{figure}[htb] 
	\centering
	\subfloat[Perception test opinion mapped to a $\beta-$PDF before and after missing detection. \label{fig:PerceptionOpinion}]{%
		\includegraphics[width=0.5\textwidth]{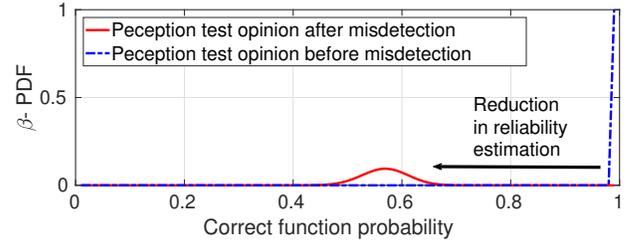}}
	\\	
	\subfloat[Effect of missing detection to reliability estimated by the perception test. The opinion is projected to the estimated reliability using $\boldsymbol{p}_{\omega} = \boldsymbol{b}_{\omega} + u_{\omega} \boldsymbol{a}_{\omega}$. At $6.8\,$s the IV's perception detects a bicycle that is not reported by the RSU. Thus the reliability estimated by the perception test is significantly decreased. \label{fig:PerceptionReliability}]{%
		\includegraphics[width=0.5\textwidth]{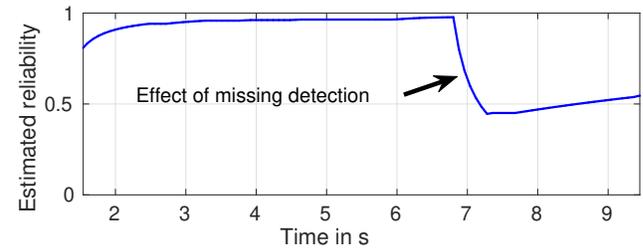}}
	\\
	\subfloat[The bicycle marked in red is not detected by the RSU.\label{fig:PerceptionVisualization}]{%
		\includegraphics[width=0.4\textwidth]{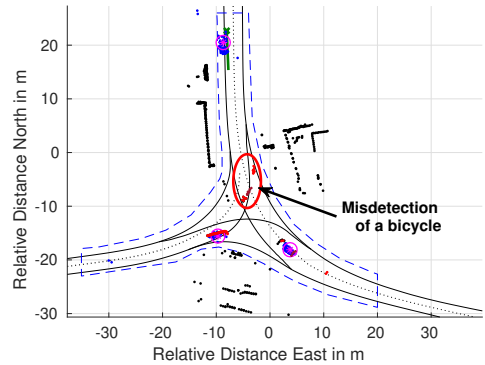}}
	\caption{Ego perception test.}
	\label{fig:PerceptionAll} 
\end{figure}

Figure \ref{fig:PerceptionAll} exemplary shows the effect of a missing detection in the RSU. First of all, the ego perception objects match the objects reported by the RSU. Hence, the estimated reliability of the RSU is high as can be seen from the blue Dirichlet PDF in Fig. \ref{fig:PerceptionAll}\subref{fig:PerceptionOpinion} and the mapped reliability in Fig. \ref{fig:PerceptionAll}\subref{fig:PerceptionReliability}. Then, the missing detection occurs as can be seen in Fig. \ref{fig:PerceptionAll}\subref{fig:PerceptionVisualization}. As consequence, the estimated reliability is significantly decreased, as can be seen in Fig. \ref{fig:PerceptionAll}\subref{fig:PerceptionReliability}. The $\beta$-PDF corresponding to the respective perception test opinion at that instance of time is depicted in red in Fig. \ref{fig:PerceptionAll}\subref{fig:PerceptionOpinion}. After the missing detections, the estimated reliability slowly recovers due to the following correctly reported information.

\subsection{Evaluation of the Comparison between Ego Localization and Received Object Lists} 

\begin{figure}[htb]
	\centering
	\includegraphics[width=0.5\textwidth]{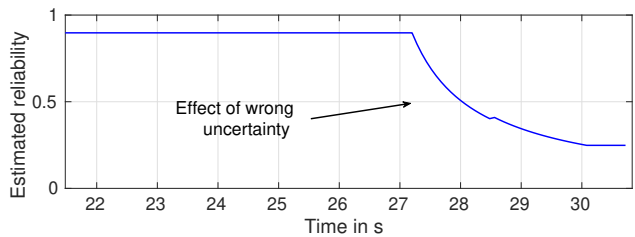}
	\caption{Localization test opinion mapped to estimated reliability using $\boldsymbol{p}_{\omega} = \boldsymbol{b}_{\omega} + u_{\omega} \boldsymbol{a}_{\omega}$. At $27.2\,$s the IV gets into the FOV of the intersection, but the position of the IV according to ego localization cannot be explained by the position and its too small uncertainty estimated by the RSU. Thus the reliability of the RSU estimated by the localization test is significantly decreased.}
	\label{fig:LocalizationAll}
\end{figure}

In Fig. \ref{fig:LocalizationAll}, the effect of an underestimated uncertainty is demonstrated by example, where the overall uncertainty $\sigma_{\text{total}}$ was chosen too small. Thus, the estimated reliability is reduced from the prior estimation, because the geometrical center of the ego vehicle according to the RTK measurement is outside the $3 \sigma$ uncertainty ellipse estimated by the RSU.

\subsection{Evaluation of the Overall Reliability Estimation}      

\begin{figure}[htb]
	\centering
	\includegraphics[width=0.5\textwidth]{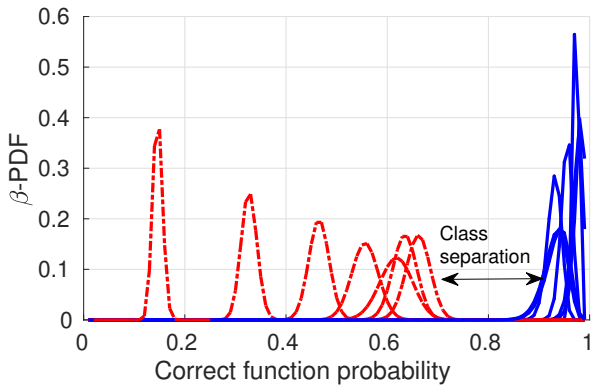}
	\caption{Overall reliability estimate. Examples of correct functionality are marked in blue, examples with detection faults are marked in dashed red.}
	\label{fig:OverallOpinion}
\end{figure}

In order to evaluate the overall reliability estimation, the reliability estimations are mapped to $\beta$-distributions and visualized all in Fig. \ref{fig:OverallOpinion}. The $\beta$-distributions resulting from the sequences with correctly performing RSU are marked in blue, while the $\beta$-distributions resulting from a faulty RSU are marked in red. It can be seen that the two classes are easily separable using the proposed reliability estimation mechanism. Furthermore it shows that the reliability is estimated to be beyond $85 \%$ whenever the RSU is reliable, while the estimated reliability is below $75 \%$ whenever the RSU is faulty. 
By integrating over the $\beta$-distributions it can be confirmed that for all true positives, a $90\%$ confidence level supports the hypothesis that the correct function probability is at least $90\%$. 
In turn, whenever the RSU is faulty, there is at least a $90\%$ confidence that the correct function probability is at most $70\%$.
This demonstrates that the reliability estimation mechanism works well and shows robustness with respect to intra-class variations.

\section{CONCLUSIONS}

In this work, a general framework to estimate the reliability of cooperative information received from an Road Side Unit (RSU) for use in intelligent vehicles (IVs) was presented. By the use of the theory of Subjective Logic (SL), we showed how orthogonal information sources (termed opinions in SL) can be fused in a natural and mathematically sound and easily extendable way. To prove the applicability, four exemplary tests were proposed, which individually only provide mild cues towards an RSU's reliability of information. However, after the proposed scheme for probabilistic fusion, IVs are able to separate faulty from correct data samples with a large margin of safety. Real-world experiments showed the applicability and effectiveness of our approach. 

In future, the possible faulty effects of wireless data communication as well as the effects of sudden and short-lived sources of error, like sensor blockage, will be examined.


\addtolength{\textheight}{-12cm}   


\bibliographystyle{IEEEtran}
{\footnotesize
	\bibliography{IV2019}}

\end{document}